\documentclass{article}%
\usepackage{amsfonts}
\usepackage{amssymb}
\usepackage{graphics}
\usepackage{graphicx}
\usepackage{makeidx}
\usepackage{latexsym}
\usepackage[dvips]{color}
\usepackage[centertags]{amsmath}
\usepackage{amsthm}
\usepackage{newlfont}
\usepackage{stmaryrd}
\usepackage{mathrsfs}
\usepackage[english]{babel}
\newcommand{\bbR}{{\mathbb R}}
\newcommand{\id}{\textrm{d}}
\begin{document}

\begin{center}
\textbf{\LARGE{Linear response in the\\ nonequilibrium zero range process}}
\end{center}
Christian Maes and Alberto Salazar\\
Instituut voor Theoretische Fysica KU Leuven, Belgium\\
{\small 
Corresponding author: A. Salazar, albertdenou@gmail.com }\\

\vspace{1cm}
\noindent \underline{Abstract}:\\
{\small 
We explore a number of explicit response formul{\ae} around the boundary driven zero range process to changes in the
exit and entrance rates. In such a nonequilibrium regime kinetic (and not only thermodynamic) aspects make a difference 
in the response. Apart from a number of formal approaches, 
we illustrate a general decomposition of the linear response into entropic and frenetic contributions, 
the latter being realized from changes in the dynamical activity at the boundaries. 
In particular in this way one obtains nonlinear modifications 
to the Green-Kubo relation.
We end by bringing some general remarks 
about the situation where that nonequilibrium response remains given by the (equilibrium) Kubo formula 
such as for the density profile in the boundary driven Lorentz gas.
}

\section{Introduction}
Linear response theory for nonequilibrium systems is slowly emerging from a great variety of formal approaches --- see \cite{njp} for a recent review. It 
remains however very important in nonequilibrium to concentrate more on the physical--operational meaning of 
the response expressions. Obviously, it is very practical to have experimental access to the various terms in a 
response formula and to learn in general to recognize facts of the unperturbed system that are responsible 
for the particular response. 
That at least is what has made the fluctuation-dissipation theorem so useful in equilibrium. 
For example, transport properties as summarised in the mobility or conductivity, can be obtained from the diffusion 
in the unperturbed equilibrium system. In other words, not only is there a unifying response relation in equilibrium, 
it also possesses a general meaning in terms of fluctuations and dissipation.  Such is not yet quite the situation for 
nonequilibrium systems and extra examples, in particular for spatially extended systems will therefore be useful.\\

The present paper gives the response systematics for the zero range process.
The zero range process regularly appears in nonequilibrium studies and has the simplifying structure that its stationary distribution is 
simple (and remains a product distribution even away from equilibrium) while it shows a rich and quite realistic phenomenology. 
We refer to \cite{shuts,EH} for a general introduction and nonequilibrium study of the model.   We refer to Section 3 in \cite{EH} for a review 
of applications, in particular for the correspondence
with shaken granular gases.  We will repeat the set-up in 
Section \ref{zrp}. Interestingly, the time-reversed zero range process has an external field and particle currents directed 
opposite to the density profile. To start however we repeat in the next section some more formal aspects of the nonequilibrium 
linear response.  
 Our point of view is to look in particular for the 
decomposition of the response into a frenetic and an entropic contribution. The entropic part is expressed in terms of 
(time-antisymmetric) currents and the frenetic part gets related to the (time-symmetric) dynamical activity.  The latter 
refers to the number of exits and entrances of particles at the boundaries of the system.
Section \ref{rzrp} performs that decomposition for the boundary driven zero range process, and gives a number of response formul{\ae} for density and current.  There we find our main results, in particular modified Green-Kubo relations.
Finally, in Section \ref{eqd} we treat some special cases which bring the nonequilibrium response to resemble 
the equilibrium Kubo formula.  That opens the separate theme of trying to understand under what physical conditions nonequilibrium 
features remain largely absent.

\section{Nonequilibrium response}\label{nonr}
We restrict ourselves to open systems  connected to various different equilibrium reservoirs.  Their nonequilibrium is passive 
in the sense that they do not affect the reservoirs directly and that all nonequilibrium forcing works directly on the 
particles of the system.   

The state of the open system is described by values $x$ for some reduced variables, e.g. (some) particle positions.  In the course of time $[0,t]$ there is a 
path or trajectory $\omega := (x_s, 0\leq s \leq t)$ for which we have a well-defined entropy flux $S(\omega)$, that is 
the change of entropy in the equilibrium reservoirs. That $S(\omega)$ typically depends on the elementary changes in $x$ and 
how that affects the energy and particle number of the reservoirs.  Of course $S(\omega)$ also depends on parameters such as 
the temperature, the chemical potentials of the reservoir and coupling coefficients of interaction.  It will thus change under 
perturbations.\\Similarly, for every state $x$ there is a notion of reactivity, like the escape rate from $x$. 
For the path $\omega$ there will then be a dynamical activity $D(\omega)$ which reflects the expected amount of changes 
along the path $\omega$, again function of the system and reservoir parameters.\\
Consider now a perturbation of that same system in which parameters are changed. Clearly, for any path $\omega$ the entropy 
flux $S$ and the dynamical activity $D$ will change.  We can look for the linear excess, that is the amount by which the perturbation 
has changed these observables to first order.  
We refer to \cite{bam09,bmw1} for the general introduction, and to \cite{mnw08} for complementary aspects to entropy.  
The linear response for a path-observable $O(\omega)$ is the difference in 
expectation $\langle \cdot\rangle^h$  between the perturbed process (with small time-dependent amplitude $h_s, s\in [0,t]$) and the 
original steady expectation $\langle\cdot\rangle$.  It has the form
\begin{equation}
\langle O(\omega)\rangle^h - \langle O(\omega)\rangle = \frac {1}{2}\langle \mbox{Ent}^{\left[ 0,t \right]}(\omega)\,O(\omega)\rangle
- \langle \mbox{Esc}^{\left[ 0,t\right] }(\omega)\,O(\omega )\rangle
\label{eq:xsusc}
\end{equation}
where Ent$^{\left[ 0,t\right] }$ is the excess in entropy flux per $k_B$ over the trajectory
due to the perturbation and Esc$^{\left[ 0,t\right] }$ 
is the excess in dynamical activity over the trajectory.
The latter and second term on the right-hand side of \eqref{eq:xsusc} is 
the frenetic contribution \footnote{the response formula \eqref{eq:xsusc} can be written in 
several equivalent ways: in the second term often there is a 
factor $1/2$ which here we include in the 
definition of the dynamical activity term in brackets; section \ref{zrp} 
will treat some specific formulations. }. 
In many nonequilibrium situations 
the physical challenge is to learn to guess 
or to find and evaluate that Esc$^{\left[ 0,t \right] }$ from partial information on the dynamics. 
The present paper takes the opportunity to explore this question and to make such task more specific for the zero range process. 
Let us however first give the more general formal structure, restricting ourselves to Markov jump processes.  
For a more general review of various recent approaches, see \cite{njp}.\\

On the finite state space $K$ we consider transition rates
$k(x,y), x,y\in K$.  We assume irreducibility so that there is exponentially fast convergence to a unique stationary
distribution $\rho(x), x\in K$, satisfying
\[
\sum_{y\in K} [\rho(x)\,k(x,y) - \rho(y)\,k(y,x)] = 0, \quad x\in
K \] 
Still, in general, there are nonzero currents of the form
$j(x,y) := \rho(x)\,k(x,y) - \rho(y)\,k(y,x)\neq 0$ for some pairs $x\neq
y\in K$, so that the stationary process is not time-reversible.\\
For physical models the rates carry a specific meaning.  Following the condition of local detailed balance the ratio
\[
\log\frac{k(x,y)}{k(y,x)} =\sigma(x,y)
\]
should be the entropy flux (in units of $k_B$) in the transition $x\rightarrow y$.  
Consider now again the path $\omega := (x_s, 0\leq s\leq t)$.  It consists of jumps (transitions) at specific times $s_i$ 
and waiting times over $s_{i+1} - s_i$.  The total entropy flux $S$ (in units of $k_B$) is
\begin{equation}\label{set}
S(\omega) = \sum_{s_i} \sigma(x_{s_i^-},x_{s_{i}}) 
\end{equation}
where the sum takes the two states of the transition $x_{s_i^-} \longrightarrow x_{s_{i}}$, with $x_{s_i^-}$ being the state 
just before the jump time $s_i$ to $x_{s_i}$.\\
For the dynamical activity we need a reference process. Writing 
\[
	k(x,y) = \psi(x,y) e^{\sigma(x,y)/2} \mbox{ with } \psi(x,y) =\psi(y,x) \mbox{ and } \sigma(x,y)=-\sigma(y,x),
\]
we take the reference rates $k_o(x,y) =1$ whenever $\psi(x,y)\neq 0$ and zero otherwise. That reference process corresponds to an infinite 
temperature limit but it will not matter 
in the end. With respect to that reference we do not only have a change 
in ``potential barrier'' $-\log 1 = 0 \rightarrow -\log \psi(x,y)$ for each transition, but also a change in the escape 
rates for each state $x$:
\[
\xi(x) = \sum_{y: \psi(x,y) > 0} [k(x,y)-1]
\]
We then take the dynamical activity $D$ over the path $\omega$  be the combination
\begin{equation}\label{setd}
D(\omega) = \int_0^t\id s \,\xi(x_s) - \sum_{s_i} \log \psi(x_{s_i^-},x_{s_{i}}).
\end{equation}
Perturbations change $S$ and $D$.  Let us look at a specific example of
perturbed transition rates considered in
\cite{diez,mprf}:
\begin{equation}\label{general}
k_s(x,y) =
k(x,y)\,e^{h_s[b V(y)-a V(x)]}, \quad t>s\geq 0
\end{equation}
where the $a,b \in \bbR$ are independent of the perturbing potential $V$ and 
the $h_s \ll 1$ is small. The coresponding perturbed Master 
equation for the 
time-dependent probability law $\rho_t$ is
\[
\frac{\id}{\id t} \rho_t(x) = \sum_y \big[k_t(y,x) \rho_t(y) - k_t(x,y) \rho_t(x)\big];
\]
while the unperturbed equations of motion are obtained by making $h_s=0$.
One standard possible choice of perturbation is
taking $a= b = 1/2T$ where $T$ is the temperature of the environment which exchanges the energy $V$ with the system. 
In general $a,b$ could be 
arbitrary; however, for the perturbed rates in \eqref{general} to satisfy 
the condition of local detailed balance, one requires that $a+b=1/T$.

We continue however with the more general perturbation \eqref{general}.
It is instructive to rewrite the perturbation \eqref{general} as
\begin{eqnarray}
k_s(x,y) &=&
k(x,y)\,e^{h_s\frac{b-a}{2}(V(x) + V(y))}\,e^{h_s \frac{a+b}{2}(V(y) - V(x))}\nonumber\\
&=& \big[\psi(x,y)e^{h_s\frac{b-a}{2}(V(x) + V(y))}\big]\;e^{\sigma(x,y)/2 + h_s (a+b)(V(y) - V(x))/2}\label{ins}
\end{eqnarray}
again being split in a symmetric prefactor (between square brackets) and an anti-symmetric part in the exponential.  
From here it is easy to see the excess for the entropy flux at a transition $x\rightarrow y$ to be
\begin{equation}\label{se}
 h_s (a+b)(V(y) - V(x))
\end{equation}
We use \eqref{se} to find the perturbation to $S$ in \eqref{set} yielding
\begin{eqnarray}\label{en}
\mbox{Ent}^{\left[ 0,t\right] }(\omega) &=& (a+b)\, \sum_{s_i} h_{s_i} [V(x_{s_{i}}) - V(x_{s_i^-}) \nonumber\\
&=& (a+b) \{h_t V(x_t) - h_0 V(x_0) - \int_0^t \id s \,\dot{h}_s V(x_s)\}.
\end{eqnarray}
For the dynamical activity we should use again the reference process with rates $k_o(x,y)$ as above.  Then,
at least for the change in escape rates (first term in \eqref{setd}) at state $x$,
\begin{eqnarray}\label{dy}
\sum_y [k_s(x,y) - k(x,y)] &=&  h_s \sum_y k(x,y)\{\frac{b-a}{2}(V(x) + V(y)) + \frac{a+b}{2}(V(y) - V(x))\}\nonumber\\
&=& h_s \,\sum_y k(x,y)[bV(y) - aV(x)]
\end{eqnarray}
to first order in $h_s$.
The total change to $D$ of \eqref{setd} is thus
\begin{equation}\label{es}
\mbox{Esc}^{\left[ 0,t\right] }(\omega) = \int_0^t\id s\, h_s\,\sum_y k(x_s,y)\,[bV(y) - aV(x_s)] + \frac{a-b}{2}\sum_{s_i} h_{s_i} [V(x_{s_{i}}) + V(x_{s_i^-})]
\end{equation}
where the last term corresponds to the change in the second term of $D$ in \eqref{setd}, as from \eqref{ins}.
In all, the expressions \eqref{en} and \eqref{es} completely specify the response \eqref{eq:xsusc} for the example \eqref{general}.\\

We can still rewrite the previous formul{\ae}, loosing somewhat the physical interpretation but gaining somewhat formal elegance.  To start, let us restrict ourselves to the more simple situation where the observable $O$  is just a state function $O(x), x\in K$.  The response then investigates the change
\[
\langle O(x_t)\rangle^h - \langle O(x_t)\rangle = \langle O(x_t)\rangle^h - \langle O\rangle
\]
to first order in the $h_s$, where the first expectation $\langle \cdot\rangle^h$ is under the perturbed Markov dynamics ($s\geq 0$) and the second $\langle \cdot \rangle$ is the original steady expectation.  To say it differently, linear response  wants to compute the generalized susceptibility $R(t,s)$ in
\[
\langle O(x_t)\rangle^h = \langle O \rangle +
\int_0^t\id s\,
h_s\, R(t,s) + o(h)\]
The nonequilibrium answer can be written in a variety of ways, many of which are rather formal, 
but they should in the end all coincide with \eqref{eq:xsusc} for \eqref{en}--\eqref{es}.  For example,
in terms of  the backward generator $L$ of the jump process,
\[
Lf(x) = \left.\frac{\id}{\id s}\right|_{s=0}\langle
f(x_s)\rangle_{x_0=x}\; = \;\sum_y k(x,y)[f(y)-f(x)]
\]
we have  for \eqref{en} that
\begin{eqnarray}
\langle  \{h_t V(x_t) - h_0 V(x_0) - \int_0^t \id s \,\dot{h}_s V(x_s)\} \,O(x_t)\rangle &=&
 \int_0^t \id s \,h_s \frac{\id}{\id s}\langle V(x_s) \,O(x_t)\rangle\nonumber\\
&=& -\int_0^t \id s \,h_s \langle V(x_s) \,LO(x_t)\rangle\nonumber
 \end{eqnarray}
On the other hand, for \eqref{es},
 \[
 \mbox{Esc}^{\left[ 0,t\right] }(\omega) = 
b\,\int_0^t\id s\, h_s\, LV(x_s) + (b-a)\{\int_0^t\id s\, h_s\, 
\sum_y k(x_s,y)\,V(x_s) - \sum_{s_i} h_{s_i} \frac{V(x_{s_{i}}) + V(x_{s_i^-})}{2}\}
 \]
We must substitute that expression together with \eqref{en} into \eqref{eq:xsusc}, which leads to
\begin{equation}\label{stationary}
 R(t,s) = a\frac{\partial}{\partial s}\langle V(x_s)O(x_t)\rangle - b\langle LV(x_s)\,O(x_t)\rangle
\end{equation}
for all times $0\leq s <t$, which recovers a result of \cite{mprf}.
 Of course, these expectations are stationary and only depend on the time-difference $t-s$.
For the case $b=a=1/(2T)$ in \eqref{general}, the response formula thus becomes
\[ 
R(t,s) = \frac{1}{2T}
\frac{\partial}{\partial s}\langle V(x_s)\,O(x_t)\rangle_{\mu}-\frac{1}{2T}\langle LV(x_s)\,O(x_t)\rangle\]
In equilibrium, i.e., under stationary time-reversal
symmetry where all currents $j(x,y)=0$, we have for $t>s$
\[
\langle LV(x_s)\,O(x_t)\rangle = \langle V(x_s)\,LO(x_t)\rangle = \frac{\partial}{\partial t}\langle V(x_s)O(x_t)\rangle 
\]
and hence the two terms in the right-hand side of
\eqref{stationary}
coincide and we recover the Kubo-formula, \cite{kubo66},
\begin{equation}\label{eqform}
R^{\mbox{eq}}(t,s) = \frac{1}{T}\,\frac{\partial}{\partial s}\left<
V(s)\,O(t)
 \right>_{\textrm{eq}},\quad 0 < s <t
 \end{equation}
  whenever $a+b = 1/T$.\\
Such formal systematics in nonequilibrium  as in \eqref{stationary} is useful as it is generally available, but its physical 
interpretation relies on the equivalence (as discussed above for the example \eqref{general}) with \eqref{eq:xsusc}, 
much in the same way as, for equilibrium, the Kubo-formula can be called a fluctuation--{\it dissipation} relation. 
Note that in nonequilibrium, from \eqref{eq:xsusc}, the response formula has become a fluctuation--{\it dissipation--activity} relation.
We refer to \cite{bam09,bmw1,mprf,njp} for more details.\\
  
Another (again more formal) possibility of writing the linear response formula \eqref{stationary} uses the adjoint $L^* $ of 
the backward generator with respect to the stationary distribution $\rho$: $\langle (Lf)\,g\rangle =\langle f\,(L^*g)\rangle$ or
\begin{equation}\label{ster}
L^*g(x) = \sum_y k(y,x) \frac{\rho(y)}{\rho(x)}\,[g(y)-g(x)]
\end{equation}
which generates the time-reversed stationary process.
 With this notation, for $s<t$ and in the stationary regime,
\begin{equation}\label{stra}
\frac{\partial}{\partial s}\langle V(x_s)\,O(x_t)\rangle = 
- \frac{\partial}{\partial t}\langle V(x_s)\,O(x_t)\rangle 
= - \langle V(x_s)\,LO(x_t)\rangle = - \langle L^*V(x_s)\,O(x_t)\rangle
\end{equation}
so that
 \[
 \frac{\partial}{\partial s}\langle V(x_s)O(x_t)\rangle + \langle LV(x_s)\,O(x_t)\rangle = 
 \langle (L-L^*)V(x_s) \,O(x_t)\rangle
 \]
Therefore, referring to the response \eqref{stationary},
\[
 a\frac{\partial}{\partial s}\langle V(x_s)O(x_t)\rangle - b\langle LV(x_s)\,O(x_t)\rangle =
(a+b)\, \frac{\partial}{\partial s}\langle V(x_s)O(x_t)\rangle - b \langle (L-L^*)V(x_s) \,O(x_t)\rangle
 \]
 or
 \begin{equation}\label{rrr}
 R(t,s) =(a+b)\, \frac{\partial}{\partial s}\langle V(x_s)O(x_t)\rangle - b \langle (L-L^*)V(x_s) \,O(x_t)\rangle
  \end{equation}
The first term is the Kubo expression \eqref{eqform}.  The second term will be useful whenever we 
know more about the time-reversed process.  If wanted, one can still substitute there
\[
((L- L^*)f)(x) = 2\sum_y\frac{j(x,y)}{\rho(x)}\,[f(y)-f(x)]
\]
to obtain the interpretation of \cite{gaw} in terms of the moving frame.\\

Let us finally mention the Agarwal-Kubo procedure for arriving at a linear response expression, see e.g. formula 13 in \cite{njp}.  
That is first order perturbation theory on the level of forward generators.  We consider the unperturbed forward generator 
\[
L^+ g(x) = \sum_y [k(y,x)g(y) - k(x,y)g(x)] 
\]
and its perturbation is denoted by ${L^h}^+$.  Then, we have in general
\begin{equation}\label{ak}
 R(t,s) =\langle \frac{({L^h}^+-L^+)\rho}{\rho}(x_s)\,O(x_t\rangle)
 \end{equation}
 The obvious disadvantage here is that one should know the stationary density $\rho$; in contrast, all observables in 
\eqref{stationary} are explicit and known and in \eqref{eq:xsusc} they even have a meaning.

\section{Formal elements of the zero range process}
\label{zrp}
\subsection{Steady state}
On the lattice interval $\{1,2,\ldots,N\}$, each site $i$ carries a number $x(i) \in \mathbb{N}$ of indistinguishable particles.  
The dynamics is characterized by the rate $w(k)$ at which a particle jumps from site $i$ when $x(i)=k$, 
and parameters $\alpha,\beta,\gamma,\delta$ for the rates of exits and entrances at the boundaries.   
More specifically, a particle moves from $i$ to a neighboring site $j=i\pm 1$ at rate $w(x(i))$. 
We need that $w(0)=0$ and $w(k)>0$ for $k>0$. At the boundary site $i=1$ a particle is added at rate $\alpha$ and at $i=N$ 
is added at rate $\delta$, while a particle moves out from $i=1$ at rate $\gamma\, w(x(1))$ and moves out from $i=N$ 
at rate $\beta\, w(x(N))$.  As reference for more details using mostly the same notation, we refer to \cite{EH}.\\
It is well-known that the product distribution $\rho = \rho_{N,\alpha,\beta,\gamma,\delta}$ is invariant,
\begin{eqnarray}\label{ste}
\rho(x) &=&  \prod_{i=1}^N \nu_i(x(i)), \quad \nu_i(k) = \frac{z_i^{k}}{{\cal Z}_i}\,\frac 1{w(1)\,w(2)\,\ldots w(k)},\; k>0\nonumber\\
{\cal Z}_i &=& 1 + \sum_{k=1}^\infty\frac{z_i^{k}}{w(1)\,w(2)\,\ldots w(k)}
\end{eqnarray}
The ``fugacities'' $z_i$ are of the form $z_i = Ci + B = z_1 + C(i-1)$ where
\[
B :=\frac{\alpha + (1-\gamma)C}{\gamma},\quad C:=\frac{\delta \gamma - \beta \alpha}{\beta \gamma N + \beta(1-\gamma) + \gamma}=
\frac{e^{\mu_r/T} - e^{\mu_\ell/T}}{N}\,\big(1 + \frac{\beta +\gamma-\beta \gamma}{\beta\gamma N}\big)^{-1} 
\]
We have introduced ``chemical potentials'' $\mu_\ell := T\log \alpha/\gamma$ and $\mu_r := T\log \delta/\beta$ with 
$T$ the environment temperature.
When $\mu_\ell=\mu_r, \alpha/\gamma = \delta/\beta$, then $C=0, B=z_i=\alpha/\gamma$; and detailed balance is satisfied.   
If not, we get a  stationary  particle current (to the right) equal to $\langle J_i\rangle = -C = \alpha -\gamma z_1 = \beta z_N - \delta$ and thermodynamic 
driving force $(\mu_\ell - \mu_r)/T = \log \alpha/\gamma - \log \delta/\beta$. Note however that for $C\neq 0$ the (then nonequilibrium) stationary 
distribution $\rho$ also depends on purely kinetic (and not only on thermodynamic) aspects; they will again enter the 
response in terms of the dynamical activity.  
For example, fixing $\alpha/\gamma$ and $\delta/\beta$ does not determine $C$, trivially but importantly.\\

\subsection{Time-reversal}
To make explicit use of the formula \eqref{rrr}, we need to know the time-reversed process, which is interesting in itself.\\ 
In general for a Markov process as we had it described in Section \ref{nonr} the time-reversed process is again a Markov jump 
process with generator $L^*$ in \eqref{ster} and with rates
\[
k^{\text rev}(x,y)  = k(y,x)\,\frac{\rho(y)}{\rho(x)}
\]
for the stationary distribution $\rho$.  Because we know the stationary distribution $\rho$ of the zero range process as the
product distribution \eqref{ste}, it is actually easy to determine explicitly the time-reversed process.  This is interesting 
also because, 
by time-reversing, the particle current will be reversed/change sign but the stationary density profile, as given in terms of the 
fugacities $z_i$, will remain the same.  As can be guessed, that only works because by time-reversing one actually generates 
an external field.  Let us see the details.\\

First we take a bulk transition in which a particle hops to a neighboring site.  
Take $y=x-e_i+e_{i+1}$ where $e_i$ stands for the particle configuration with exactly one particle at site $i$.  Then,
\[
\frac{\rho(y)}{\rho(x)} = \frac{z_{i+1}}{z_i}\frac{w(x(i))}{w(x(i+1))},\quad k(y,x) = w(x(i+1)) 
\]
which means that in the time-reversed process a particle moves from site $i$ to $i+1$ at rate 
$k^{\text rev}(x,x-e_i+e_{i+1})  = z_{i+1} \,w(x(i))/z_{i}$ while 
similarly for a jump from $i$ to $i-1$, $k^{\text rev}(x,x-e_i+e_{i-1})  =  z_{i-1} \,w(x(i))/z_{i}$.  
We have therefore for the time-reversed process again a zero range process but now in an inhomogeneous bulk field 
\[
E_i := 2\log \frac{z_{i+1}}{z_{i}}
\]
over the bond $(i,i+1)$, having the sign of $C$, i.e., pushing the particles towards the boundary {\it where the chemical potential 
was largest}. 
At the boundaries we find the creation and annihilation parameters for the time-reversed process to be
\[
\alpha^{\text rev} = 
\gamma z_1,\quad \beta^{\text rev} = \frac{\delta}{z_N},\quad \gamma^{\text rev}= \frac{\alpha}{z_1},\quad \delta^{\text rev} =\beta z_N.
\]
That means that the chemical potentials for the reversed process have become 
\begin{eqnarray}
\mu_\ell^{\text rev} &=& - \mu_\ell +2 T\log (e^{\mu_\ell/T} + C/\gamma)\nonumber\\
\mu_r^{\text rev} &=& -\mu_r + 2T \log(e^{\mu_r/T} - C/\beta)\nonumber
\end{eqnarray}

Note of course that in the case of detailed balance $E_i\equiv 0 $ and $\alpha^{\text rev} =\alpha$ etc., 
so that the equilibrium process is unchanged
by time-reversal.\\

We can now write down the explicit expression for the second term in \eqref{rrr}:
\begin{eqnarray}
(L - L^*)V\,(x) &=& \big(\gamma-\frac{\alpha}{z_1}\big)\,w(x(1))\,[V(x-e_1)-V(x)] + \big(\alpha-\gamma z_1\big)\,[V(x+e_1)-V(x)] \nonumber\\
&+& \big(\beta-\frac{\delta}{z_N}\big)\, w(x(N))\,[V(x-e_N)-V(x)] + \big(\delta -\beta z_N\big)\,[V(x+e_N)-V(x)]\nonumber\\
&&-C\sum_{i=1}^{N-1} \frac{w(x(i))}{z_i} [V(x-e_i+e_{i+1})-V(x)]\nonumber\\
&&+C\sum_{i=2}^{N} \frac{w(x(i))}{z_i} [V(x-e_i+e_{i-1})-V(x)]\nonumber
\end{eqnarray}
Applying that for $V(x) = {\cal N}(x):= x(1) + x(2) +\ldots + x(N)$ the total number of particles in the system, we get
\begin{equation}\label{lminl}
(L - L^*){\cal N}\,(x) =\big(\frac{\alpha}{z_1}-\gamma\big)\,w(x(1)) +  \big(\frac{\delta}{z_N} - \beta\big)\, w(x(N))
\end{equation}
where we have also used that $\gamma z_1 + \beta z_N = \alpha + \delta$.

\section{Responses in the zero range process}\label{rzrp}
Let us consider the perturbation
\begin{equation}\label{mape}
\alpha \rightarrow q\,\alpha,\quad \beta\rightarrow p'\,\beta,\quad \gamma \rightarrow p\,\gamma,\quad \delta\rightarrow q'\,\delta
\end{equation}
to the parameters governing the entrance and exit rates at the boundaries of the system.
Their thermodynamic meaning is to shift the chemical potentials by $h_\ell = T\log q/p$ for the left and by 
$h_r = T\log q'/p'$ for the right reservoir.   Depending on the remaining freedom how to choose the $p, p'$ we can distinguish still several ``kinetic'' possibilities.

\subsection{``Potential'' perturbation}\label{potp}
A first possible perturbation that we consider is that
\begin{equation}\label{shi}
\frac{q}{p} = \frac{q'}{p'} = e^{ h/T}
\end{equation}
with $h$ the small (equal) shift in left and right chemical potential.
Even while the zero range process is not formulated directly in terms of a potential, even at detailed balance, 
it is still easy to fit \eqref{shi} into the scheme of \eqref{general}, in particular by choosing 
$h_t\equiv h$ (time-independent),  $a=b=1/(2T)$, potential $V={\cal N}$ equal to the particle number, and
\begin{equation}\label{potper}
e^{h/(2T)}=q=q', \quad e^{-h/(2T)}=p=p'.
\end{equation}
  We can thus apply \eqref{rrr} with  formula \eqref{lminl} to give the correct modification of the Kubo formula as 
 \begin{eqnarray}\label{rrrtotc}
\frac{\langle O(x_t)\rangle^{h}- \langle O \rangle}{h} &=& \frac 1{T} \langle {\cal N}\, O\rangle - \frac 1{T}\,\langle {\cal N}(x_0)\, O(x_t)\rangle \nonumber\\
&& +\frac 1{2T}\int_0^t\id s\{\big(\frac{\alpha}{z_1}-\gamma\big)\,\langle w(x_0(1))\,O(x_s)\rangle + \big(\frac{\delta}{z_N}-\beta\big)\, \langle w(x_0(N))\,O(x_s)\rangle\}  
  \end{eqnarray}
Of course we could also have used \eqref{stationary} with $L{\cal N}(x) = \alpha + \delta - \gamma w(x(1)) -\beta w(x(N))$ to obtain
\begin{eqnarray}\label{rpo}
\frac{\langle O(x_t)\rangle^{h}- \langle O \rangle}{h} &=& \frac 1{2T} \langle {\cal N}(x_t) - {\cal N}(x_0); O(x_t)\rangle \nonumber\\
&& +\frac{1}{2T}
\int_0^t\id s\{\gamma\,\langle w(x_0(1));O(x_s)\rangle + \beta\, \langle w(x_0(N));O(x_s)\rangle\}  
  \end{eqnarray}
where we have used connected correlation functions $\langle A;B\rangle := \langle A\,B\rangle  - \langle A\rangle\,\langle B\rangle$. The first term in the right-hand side 
 is the entropic or dissipative part of the response, since in that correlation
one sees the observable $O$ correlated with the particle loss; the last term may be called the frenetic part of the response, since one meets there the correlation with the time-integrated escape rates.\\

Finally one finds place for the Agarwal-Kubo formula \eqref{ak}, which here is explicit because the stationary density 
$\rho$ is given in \eqref{ste}.  For the ``potential'' perturbation \eqref{shi}--\eqref{potper} given 
by $\alpha \rightarrow (1+ h/(2T)) \alpha, \beta \rightarrow (1- h/(2T)) \beta, \gamma \rightarrow (1- h/(2T)) \gamma, \delta \rightarrow (1+ h/(2T)) \delta$ and under discussion so far, that gives
\begin{eqnarray}\label{ak12}
\frac{{L^h}^+\rho - L^+\rho}{\rho}(x) 
=  \alpha \frac{h}{2T}\, [\frac{\rho(x-e_1)}{\rho(x)} - 1] + \delta \frac{h}{2T}\,[\frac{\rho(x-e_N)}{\rho(x)} -1]
+ \gamma\frac{h}{2T}\,w(x(1)) \nonumber\\
-  \gamma\frac{h}{2T}w(x(1)+1)\frac{\rho(x+e_1)}{\rho(x)} +\beta\frac{h}{2T}\,w(x(N)) 
-  \beta\frac{h}{2T}w(x(N)+1)\frac{\rho(x+e_N)}{\rho(x)}\nonumber\\
= \frac{h}{2T}\{\frac{\alpha}{z_1}(w(x(1)) - z_1) + \frac{\delta}{z_N}(w(x(N)) - z_N)
+\gamma( w(x(1)) - z_1)  + \beta (w(x(N)) - z_N) \}
\end{eqnarray}
This calculation results in the linear response formula
\begin{equation}\label{lak}
\frac{\langle O(x_t)\rangle^{h}- \langle O \rangle}{h} = \frac 1{2T}
\int_0^t\id s\{\big(\frac{\alpha}{z_1}+ \gamma\big)\,\langle w(x_0(1));O(x_s)\rangle 
+ \big(\frac{\delta}{z_N}+ \beta\big)\, \langle w(x_0(N));O(x_s)\rangle\}  
\end{equation}

\subsection{General perturbation}

We emphasize that the three response formul{\ae} \eqref{rrrtotc}--\eqref{rpo}--\eqref{lak} are mathematically identical. 
They all start from the ``potential perturbation'' \eqref{general} as realized in \eqref{shi}--\eqref{potper}. They are 
however not to be applied for other perturbations even consistent with \eqref{shi}, except in equilibrium where the 
response does not pick up the detailed kinetics.
Let us therefore do better (more general) and  illustrate the systematic interpretation  with unique formula \eqref{eq:xsusc} to 
the perturbation \eqref{mape}.\\

We only need experience with entropy and no calculation to find the first term in \eqref{eq:xsusc}.  For the perturbation \eqref{mape} 
the entropic part in the response follows the usual (irreversible) thermodynamics and we must have the excess in entropy flux given by
\begin{equation}\label{fer}
 \mbox{Ent}^{\left[ 0,t \right]}(\omega) =  -\frac{h_r}{T}\,J_r(\omega) -\frac{h_\ell}{T}\,J_\ell(\omega)
\end{equation}
where $J_r$ ($J_\ell$) is the net number of particles that have exited to the right (left) reservoir (time-integrated current).
When we specify to a perturbation like \eqref{shi} in which the chemical potentials get shifted together, $h=h_r=h_\ell$, we can use that $J_\ell(\omega) + J_r(\omega) = {\cal N}(x_0) - {\cal N}(x_t)$ so that the excess in entropy flux becomes
\begin{equation}\label{gfer}
 \mbox{Ent}^{\left[ 0,t \right]}(\omega) =  \frac{h}{T}\,({\cal N}_t - {\cal N}_0)
\end{equation}
 proportional to the change over time in particle number.\\
 For the second term in \eqref{eq:xsusc} we lack the experience and calculation will guide us.  The point is that the 
dynamical activity \eqref{setd} exactly picks up the time-symmetric part in the action for path-integration. More specifically, 
let us now call $P^h$ the process started from the unperturbed stationary zero range process \eqref{ste} but under the 
perturbed dynamics for a time $[0,t]$.  The unperturbed stationary process is denoted by $P$.  We can compute the action 
${\cal A^h}$ for which
\[
P^h = e^{-{\cal A}^h}\,P \simeq (1-{\cal A}^h)\,P
\]
with
\begin{eqnarray}
{\cal A}^h &=& - I^\ell_{\shortleftarrow}\,\log p- I^\ell_{\shortrightarrow}\,\log q  - I^r_{\shortrightarrow}\,\log p' -  
I^r_{\shortleftarrow}\,\log q'\nonumber\\
  &+& \int_0^t \id s \{(p-1)\,\gamma\, w(x_s(1)) + (p'-1)\,\beta\, w(x_s(N)) + (q-1)\alpha + (q'-1) \delta \}
\end{eqnarray}
where for example $I^\ell_{\shortrightarrow}$ equals the total number of particles that have entered the system from 
the left, and $I^r_{\shortrightarrow}$ is the total number of particles that have escaped to the right reservoir.
We decompose this action with the time-reversal $\theta$ which makes $(\theta x)_s= x_{t-s}$, so that
the response (up to higher order in $h$) can be obtained from
\begin{eqnarray}
P^h - P &=& \frac 1{2}[{\cal A}^h\theta - {\cal A}^h]\,P - \frac 1{2}[{\cal A}^h\theta + {\cal A}^h]\,P\nonumber\\
&=& \{\frac 1{2} \mbox{Ent}^{\left[ 0,t \right]}
-  \mbox{Esc}^{\left[ 0,t\right] }\}\,P
\end{eqnarray}
where we indicate the general relation with \eqref{eq:xsusc}.\\
In particular, we verify that
\[
{\cal A}^h\theta - {\cal A}^h = \log\frac{q}{p} \,(I^\ell_{\shortrightarrow} - I^\ell_{\shortleftarrow} ) +
\log\frac{p'}{q'} \,(I^r_{\shortrightarrow} - I^r_{\shortleftarrow} ) 
\]
indeed exactly equals \eqref{fer} (using for example $I^\ell_{\shortrightarrow} - I^\ell_{\shortleftarrow} = -J_\ell$). 
On the other hand, for the time-symmetric part
\begin{eqnarray}
{\cal A}^h\theta + {\cal A}^h &= & -\log (pq) \,I^\ell - \log (p'q') \,I^r 
+ 2(p-1) \gamma \int_0^t \id s \,w(x_s(1))\nonumber\\ &+& 2(p'-1) 
\beta \int_0^t \id s\, w(x_s(N)) + 2(q-1) \alpha  t + 2 (q'-1) \delta \,t
\end{eqnarray}
with left activity $I^\ell := I^\ell_{\shortleftarrow} + I^\ell_{\shortrightarrow}$ the total number of 
transitions at the left boundary and similarly for $I^r$ at site $N$.  
The excess in dynamical activity $\mbox{Esc}^{\left[ 0,t\right] } = ({\cal A}^h\theta + {\cal A}^h)/2$ that we need for 
the general response in \eqref{eq:xsusc} is thus
\begin{eqnarray}\label{fes}
\mbox{Esc}^{\left[ 0,t\right] }(\omega) &= & -\log \sqrt{pq} \,I^\ell - \log \sqrt{p'q'} \,I^r 
+ (p-1) \gamma \int_0^t \id s \,w(x_s(1))\nonumber\\ &+& (p'-1) \beta \int_0^t \id s\, w(x_s(N)) + (q-1) \alpha  t +  (q'-1) \delta \,t
\end{eqnarray}
Note that of course here the separate $p,p'$ and $q,q'$ play a role, and not just their ratio
$p/q, p'/q'$ as for \eqref{fer} --- that is how the frenetic contribution picks up kinetic information, while 
the entropic part is purely thermodynamic.
Substituting \eqref{fer} and \eqref{fes} into \eqref{eq:xsusc} gives the general response of the zero range process under \eqref{mape}.
A natural application is to look at how the current into the left reservoir changes when $h_r=0, h_\ell = -a$ or $q'=p=p'=1$ but $q=1 -a/T$, 
decreasing (for $a>0$) the chemical potential of the left reservoir.  Then, for that choice, \eqref{fer} and \eqref{fes} give
\begin{equation}\label{gkr}
\langle J_\ell\rangle^h - \langle{J_\ell}\rangle =  \frac{a}{2T}\langle J_\ell;J_\ell\rangle -\frac{a}{2T}\langle J_\ell;I^\ell\rangle
\end{equation}
which is the modification to the Green-Kubo relation \cite{yan10} , for all times $t>0$, for the boundary driven zero range process. 
Observe that it is the correlation between current $J_\ell$ and dynamical activity $I^\ell$ that governs the correction. 
When $t\uparrow +\infty$, the conductivity will of course coincide with the change of $C$ in \eqref{ste} under $\alpha$.
There is a similar relation for the change in expected dynamical activity, so that in fact
\[
\langle J_\ell + I^\ell \rangle^h - \langle{J_\ell + I^\ell}\rangle =  
\frac{a}{2T}\langle J_\ell;J_\ell\rangle -\frac{a}{2T}\langle I^\ell;I^\ell\rangle
\]
is given by a difference between variances of the current and dynamical activity,
where still $\langle J_\ell\rangle = C =-\alpha+\gamma z_1,\langle I^\ell\rangle = \alpha + \gamma z_1$.\\
Formul{\ae} \eqref{gfer}--\eqref{fes} in \eqref{eq:xsusc} will of course also lead again to a formula equal 
to each of the \eqref{rrrtotc}--\eqref{rpo}--\eqref{lak} when restricting to \eqref{shi}--\eqref{potper}.

\subsection{``External'' perturbation} 
Shifting the chemical potentials (from the outside) realistically means to change $\alpha\rightarrow q\,\alpha$ and 
$\delta \rightarrow q'\, \delta$ but not the exit rates $\beta$ and $\gamma$.  That is thermodynamically the same 
(in the shift of chemical potentials) as for the ``potential'' perturbation in Section \ref{potp} but it is kinetically different. 
The response formul{\ae} \eqref{rrrtotc}--\eqref{rpo}--\eqref{lak} are then invalid except at equilibrium.
 Here we look when we change  only the rates of the incoming particles in \eqref{mape} but restricting ourselves to \eqref{shi}:
\begin{equation}
\label{rep}
p=1=p',\quad q=q'=1+ h/T
\end{equation}
  Note that the expected total activity in the unperturbed steady regime equals
\[
\langle I^\ell+I^r \rangle = (\alpha + \gamma z_1 + \beta z_N + \delta)t = 2(\alpha + \delta)t
\]
because the stationary current equals $\alpha-\gamma z_1 = \beta z_N - \delta$.
That means that the excess dynamical activity \eqref{fes} (for perturbation \eqref{rep}) simply equals
\begin{equation}
\mbox{Esc}^{\left[ 0,t\right] }(\omega) =  \frac{h}{2T} \{ \langle I^\ell+I^r \rangle  - [I^\ell + I^r]\} 
\end{equation}
which is now very visibly related to the dynamical activity.  We therefore find the linear response formula  \eqref{eq:xsusc} to become
\begin{equation}\label{LINR}
\frac{\langle O(\omega)\rangle^h - \langle O(\omega)\rangle}{h} = \frac{1}{2T} \,\langle({\cal N}_t - {\cal N}_0);O(\omega)\rangle 
+\frac{1}{2T}\langle (I^\ell+I^r); O(x)\rangle
\end{equation}
which is another result for the linear response of the boundary driven zero range model when both left and right entrance rates have been 
increased with the same small amount. Note that from \eqref{eq:xsusc} it is here also possible to take a general path-observable 
$O(\omega)$ that depends on the whole trajectory $\omega$.  The first term is entropic corresponding to the dissipation of particles and 
the second term is frenetic with the total dynamical activity 
$I := I^\ell+I^r = I^\ell_{\shortleftarrow} + I^\ell_{\shortrightarrow} +I^r_{\shortleftarrow} + I^r_{\shortrightarrow}$.\\  
Let us check the formula \eqref{LINR} for the linear response around equilibrium ($C=0$, detailed balance), 
and with $O = I$ the total activity.  Then, since the first term $\langle({\cal N}_t - {\cal N}_0);O(\omega)\rangle^{\text eq} = 0$ 
for time-symmetric $O$, we have a Green-Kubo type formula for the linear response of the dynamical activity around equilibrium:
\begin{equation}
\frac{\langle I \rangle^h - \langle I\rangle^{\text eq}}{h} =\frac{1}{2T}\,\mbox{Var} I > 0
\end{equation}
with, in the right-hand side, the unperturbed equilibrium variance of the dynamical activity giving the expected change in 
that same dynamical activity when the left and right chemical potentials get slightly shifted.  Whether, say for positive $h$, 
the change in dynamical activity remains positive also for boundary driven zero range processes depends apparently on whether 
the dynamical activity is positively or negatively correlated with the dissipation of particles.  One could guess that 
for very small $\alpha, \delta \ll 1$ while keeping $ \gamma,\beta w(k) \simeq 1$ (low temperature reservoirs) there is 
a negative correlation between ${\cal N}_t - {\cal N}_0$ and $I$ which would make at least the first 
term in \eqref{LINR} for $O=I$ negative.\\
In any event however, be it equilibrium or nonequilibrium, we have the positivity of
\begin{equation}
 \frac{\langle {\cal N}(x_t) + I\rangle^h - \langle {\cal N}(x_0)+ I\rangle}{h} =\frac{1}{2T} \,\,\mbox{Var} ({\cal N}_t - {\cal N}_0 + I)  
> 0
\end{equation}
by taking the observable $O =  {\cal N}_t - {\cal N}_0 + I$ in \eqref{LINR}.\\

Let us further simplify and take $O$ in \eqref{LINR} a state function. It is then relevant to see how the stationary 
distribution \eqref{ste} gets modified under \eqref{rep}.  It is straightforward to check that $C, B \rightarrow qC, qB$ so that 
the new ``fugacities'' become equal to $qz_i$.  The stationary distribution thus simply changes by multiplying $\exp[h{\cal N}(x)/T]$ 
to the weights $\rho(x)$. It is therefore not so surprising that the linear response drastically simplifies.  
To check it we take the opportunity to illustrate again the Agarwal-Kubo procedure \eqref{ak} but now for the perturbation \eqref{rep}:
\begin{eqnarray}
\frac{{L^h}^+\rho - L^+\rho}{\rho}(x) &=&   
\alpha (q-1)\, [\frac{\rho(x-e_1)}{\rho(x)} - 1] + \delta (q'-1) \,[\frac{\rho(x-e_N)}{\rho(x)} -1]\nonumber\\
&=& \alpha \frac{h}{T}\, [\frac{w(x(1))}{z_1} - 1] + \delta  \frac{h}{T} \,[\frac{w(x(N))}{z_N} -1]\nonumber
\end{eqnarray}
where we substituted the known stationary distribution $\rho$ from \eqref{ste}.  On the other hand, the backward generator 
of the time-reversed process equals
\[
L^*{\cal N}\,(x) =  -\frac{\alpha}{z_1} \,w(x(1)) + \gamma z_1 + \beta z_N  -\frac{\delta}{z_N}\,w(x(N)) 
\]
and $\alpha-\gamma z_1 + \delta -\beta z_N=0$.
Therefore,
\begin{equation}\label{acc}
\frac{{L^h}^+\rho - L^+\rho}{\rho} = -\frac{h}{T}\,L^* {\cal N}
\end{equation}
As a consequence, using \eqref{ak} results in the linear response exactly of the same form \eqref{eqform} as in equilibrium, because
 (with $V = {\cal N}$ in \eqref{stra}),
 \begin{eqnarray}
\frac{\id}{\id s} \langle {\cal N}(x_s)\,O(x_t)\rangle &=& - \frac{\id}{\id t} \langle {\cal N}(x_0)\,O(x_{t-s})\rangle\nonumber\\
&=& -\langle {\cal N}(x_0) LO(x_{t-s})\rangle = -\langle L^*{\cal N}(x_0)\,O(x_{t-s})\rangle
\label{kuboform}
\end{eqnarray} 
In other words, for state observables the linear response of any boundary driven zero range process 
to ``external'' perturbations \eqref{rep} 
has always the same equilibrium Kubo-form \eqref{eqform}, independent of being close or far from detailed balance.

\section{Intersections of equilibrium and nonequilibrium evolutions}\label{eqd}
The difference between equilibrium and nonequilibrium processes is not always so crystal clear.  For exampe, if one starts 
with a dynamics for which the Gibbs distribution $\sim e^{-\beta H}$ is invariant, for some Hamiltonian $H$, then that 
distribution is also obviously unchanged when adding extra transformations or updating that leave the Hamiltonian $H$ invariant.  
On a more formal level, suppose we modify the Liouville equation to
\begin{equation}
\frac{\partial}{\partial t} \rho(x,t) + \{\rho,H\} = \int\id x [k(y,x)\,\rho(y) - k(x,y)\, \rho(x)]
\end{equation}
where the right-hand side involves transition rates $k(x,y)$ between states $x\rightarrow y$.  
If these $k(x,y)$ are zero unless $H(x) = H(y)$, then $\rho\sim \exp[-\beta H]$ remains of course invariant.  
On the other hand, the modified dynamics need not at all to satisfy detailed balance and then the resulting 
stationary regime will not be time-reversal invariant.

The Kubo formula \eqref{eqform} summarizes equilibrium linear response in terms of a fluctuation-dissipation formula.  As we have seen in the previous 
section with the combination 
\eqref{acc}--\eqref{kuboform}, 
the Kubo formula extends to the zero range process and for 
external perturbations \eqref{rep} to the nonequilibrium case.  
In the present section we look at that from a more general perspective.

\subsection{Special perturbations}\label{spp}
A special case arises when $b=0$ and $a=1/T$ in (\ref{general}),
because then the response is of the equilibrium form \eqref{eqform}.\\
Suppose we have (quite arbitrary) a Markov jump process with rates $k(x,y)$ that we perturb
by adding a time-dependent potential into
\begin{equation}\label{suft}
k_t(x,y) = k(x,y)\, e^{-h_t V(x)/T}
\end{equation}
where $h_t$ is the small parameter.  
The linear response formula is obtained by putting $b=0$ in \eqref{stationary} 
which gives the Kubo-equilibrium formula.\\
That can also be seen
from the following consideration. Take $h$ to be constant; the law
$\rho^h$ defined by $\rho^h(x) \propto \rho(x)e^{ h V(x)/T}$ is
stationary for the new dynamics (to all orders in $h$).  In other
words, here the resulting behavior under this perturbation is like
in equilibrium, even though the unperturbed dynamics can be far
from equilibrium.\\

The case of perturbation \eqref{rep} for zero range is just slightly different and is summarized in \eqref{acc}, which 
is the condition that there exists a function $V$ for which
\[
({L^h}^+ - L^+)\rho = h\,\rho\,L^*V = h\, L^+(V\rho)
\]
 for the stationary density $\rho$.  That
is equivalent with  finding a potential $V$ so that for all functions $f$
\begin{equation}\label{suf}
\sum_x ((L^h-L)f )(x)\,\rho(x) = h\sum_x (Lf)(x)\, V(x)\, \rho(x)
\end{equation}
It is easily seen that \eqref{suf} exactly follows when $L^h = (1+ hV)\,L$ which (basically) is 
\eqref{suft}.  Therefore, \eqref{acc} or \eqref{suf} is only slightly weaker than \eqref{suft}.

\subsection{Density response in the boundary driven Lorentz gas}\label{lg}
\begin{figure}\begin{center}\includegraphics[scale=0.9]{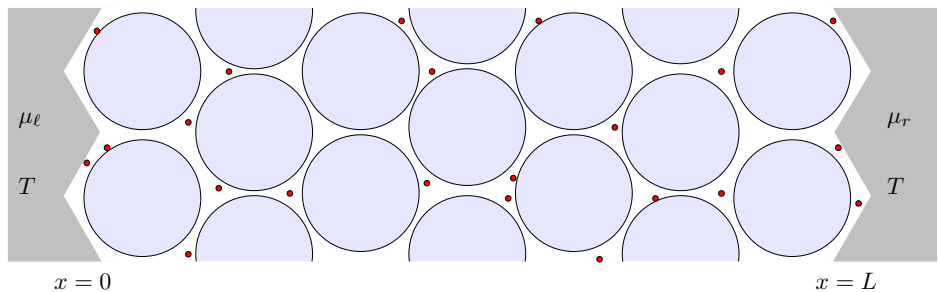}\caption{The boundary driven Lorentz gas. 
A flat rectangular slab is placed between two thermo-chemical reservoirs and contains 
an array of fixed discs, which scatter particles (red dots) via elastic collisions. 
The centers of the scatterers of radius $R$ are placed in a regular triangular lattice with {\it finite horizon}; that is, 
the distance among the centers
of contiguous disks ($4R/\sqrt 3$) ensures that a particle cannot cross the distance of a unit 
cell without 
colliding at least once with a scatterer. There is a uniform temperature in the 
reservoirs $T$, which determines the velocities of all gas particles. 
In the molecualr dynamics simulation, when a particle hits a boundary wall 
it disappears from the system, while other particles are injected 
to the system at given rates, proportional to each reservoir density.
}\label{fig:lgs}\end{center}\end{figure}
The Lorentz gas is a well known mechanical model of particle scattering that reproduces 
electron transport in metals \cite{lor05, dru00}.  Concerning our present focus and subject what becomes 
important is the fact that in the appropriate scales of time and energy the 
Lorentz gas is diffusive, see \cite{sza00, kla07} and references therein. 
Moreover, when the system is connected to reservoirs, 
the ``external'' perturbations \eqref{rep} become very natural. 
Thus, one can expect that the response for the density profile follows the zero range process 
as studied in the previous sections.  
We have performed 
extensive numerical experiments in such model to corroborate our expectations. \\

To be more precise, consider the two-dimensional slab containing a Lorentz gas illustrated in Fig \ref{fig:lgs}. 
There is a cloud of point particles which 
move freely in the space between the array of scatterers and collide elastically with them. 
The vertical coordinate is periodic and in the horizontal direction 
there are left and right boundary walls, which connect the system to thermo-chemical reservoirs, 
characterized by chemical potentials $\mu_\ell, \mu_r$ with uniform 
temperature $T$. In terms of the mean reservoir density $\rho $, the reservoir chemical potential 
$\mu \propto T\ln (\rho/T)$. During time evolution, as a particle hits the boundaries, it moves into a reservoir; 
additionally, other particles are emitted to the system at given rates 
$\pi _{\ell,r}\sim \rho _{\ell,r}\sqrt{T}$ and incoming velocities taken 
from Maxwellians at temperature $T$. 
The complete model of stochastic thermal and particle reservoirs connected to 
the Lorentz slab is 
borrowed from a similar work on a modified Lorentz gas; a detailed description about the choice 
of emission rates and chemical potential, temperature and incoming particle velocities from the reservoirs 
can be found there \cite{splg03}. In our present case we are interested in independent particles 
with constant temperature $T$; with this setting in mind the planar Lorentz gas slab of 
Fig. \ref{fig:lgs} evolves to a nonequilibrium stationary state with diffusive transport of particles, 
whenever $\Delta \mu \equiv \mu_\ell - \mu_r\neq 0$.\\

We now wish to connect this model with the zero range model. 
The rates at which particles enter (like $\alpha$ and $\delta$ in 
the zero range process) are controlled externally by the nominal reservoir 
temperature and the chemical potentials. 
For the rates at which individual particles leave, that is only controlled by the temperature and the local 
(boundary) density. Thus, one is under perturbation \eqref{rep}.  We have therefore proceeded to test 
whether our boundary driven Lorentz gas satisfies the response as predicted by the Kubo-formula \eqref{eqform} 
independent of the distance to equilibrium.  The simulation result is indeed positive. 

\begin{center}\begin{figure}\includegraphics[scale=1.2]{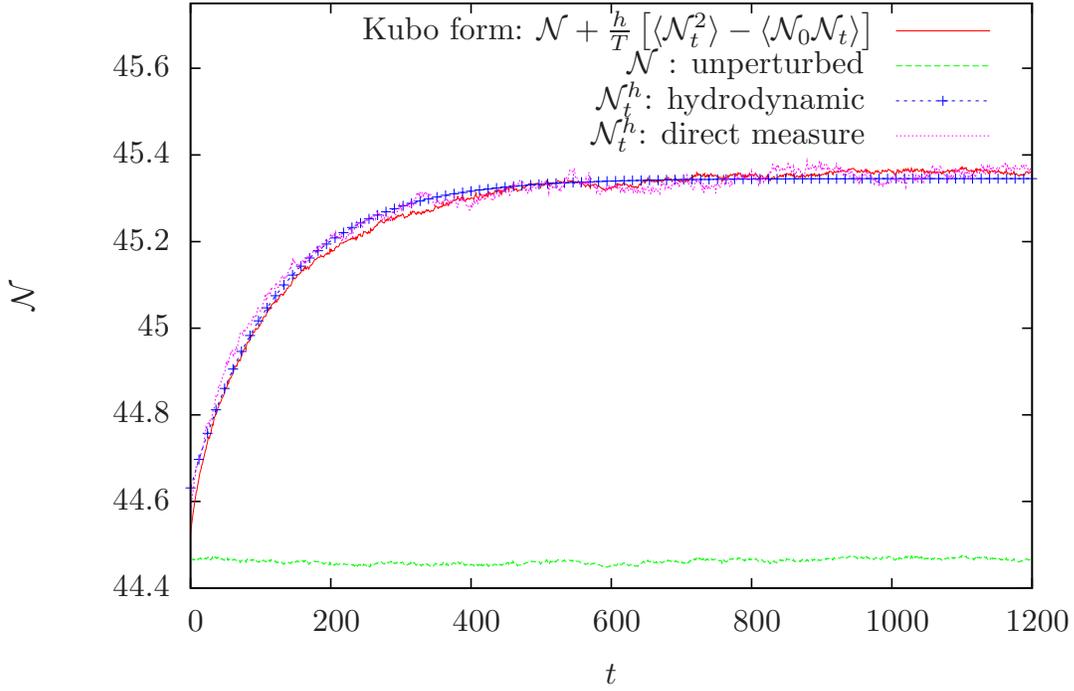}
\caption{The response in the number of particles 
$\mathcal{N}$ of the driven Lorentz gas when both reservoir chemical potentials are shifted, 
$\mu_{\ell,r}\rightarrow \mu_{\ell,r} + h$. 
The full curve is the Kubo-equilibrium formula, calculated with $\Delta \mu /T=0.2$, $T=150$. 
The dotted curve 
corresponds to direct measurements of $\mathcal{N}$ while performing the shift at $t=0$. These curves are 
obtained from averages over $1.5\times 10^6$ initial conditions. Also in the plot, 
the crosses (blue) show the response obtained by solving 
the diffusion equation $\partial _t \rho(x,t)= \lambda \partial _{xx}\rho(x,t)$, with $\lambda $ diffusivity, 
taking the stationary unperturbed particle density profile as initial condition, and perturbed densities 
as boundary conditions. 
}\label{fig:responselgs}
\end{figure}\end{center}

We have carried out nonequilibrium molecular dynamics simulations of the system in Fig. \ref{fig:lgs} and have taken 
as observable the total number ${\cal N}$ of particles in the system. 
The perturbation simply consists of modifying the reservoir densities, so that the entrance rates $\pi _{\ell,r}$ 
are shifted by the same small amount $\pi _{\ell,r} \rightarrow \pi _{\ell,r}e^{h/T}$ (depending also on the constant 
temperature). The response of $\mathcal{N}_t$ to this perturbation is shown in Fig \ref{fig:responselgs} 
for a nonequilibrium stationary regime with moderate driving of $\Delta \mu/T=0.2$ and $T=150$, which relaxes to 
a new stationary regime with different chemical potentials. The perturbation 
is applied at time $t=0$ and the system is then observed in transient states which evolve to the new stationary state. 
Each response curve consists of averages over an ensemble of $1.5\times 10^3$ initial conditions from the steady regime; 
relaxation to the final (stationary) state takes about $9.15\times 10^4$ collisions in the gas. 
The response for a similar setting with a higher driving $\Delta \mu /T = 2.0$, and using either 
of the terms in \eqref{kuboform}, gives similar outcomes: indeed we see that 
the Kubo-relation \eqref{eqform} follows no matter how far from equilibrium we are. That is not surprising because of the 
independence of the particles; actually we can predict all density 
responses simply from solving the linear diffusion equation. This is also shown in 
Fig. \ref{fig:responselgs} with the curve in crosses. Yet, one must note that this interesting example 
is just a special case of what happens more generally 
in the zero range model (possibly showing non-linear hydrodynamics).

\section{Conclusions}
One of the less understood facts of nonequilibrium physics is that the regime of linear response around equilibrium appears 
to extend sometimes quite beyond its theoretical boundaries.  Depending on the situation, that is the case for certain 
transport equations like the Fourier or even sometimes Ohm's  law, but also for the more general regime of hydrodynamics where local 
equilibrium often appears to be a very good approximation.  In nonequilibrium and irreversible thermodynamics, 
Green-Kubo relations and general principles like the minimum/maximum 
entropy production principle often continue to work 
and are used beyond their theoretical limits of validity.  

In fact, one of the reasons for not having yet an established nonequilibrium statistical mechanics may well be the lack of 
urgent questions as irreversible thermodynamics continues to work surprisingly well in a large range of transport and 
rate processes in physical or chemical systems.  Much of standard thermodynamics can even be mimicked for relatively 
small systems without feeling the urge for new concepts beyond those available in close-to-equilibrium regimes.  
Only with turbulence and very-far-from-equilibrium processes where new phenomena such as pattern formation and 
self-organization appear, do we really see major modifications with respect to the traditional approach. \\

In this paper we have studied response in the nonequilibrium zero range process, 
giving explicit expressions of the entropic and frenetic terms in which such response is formally decomposed.
That was done for various types of perturbations to the boundary rates.  We have found systematic contributions
of correlation functions with the dynamical activity to correct in general the Kubo-equilibrium formula.
There are in particular modified Green-Kubo relations where the current and the dynamical activity complement their responses.
There is however also an important case of ``external'' perturbations where the response retains the equilibrium form; 
that can also be checked for the driven Lorentz gas, which is a microscopic mechanical model.
We may expect similar behavior for other boundary driven systems with diffusive transport 
for which the analogy with certain aspects of the zero range process can be argued. \\


\begin{thebibliography}{99}                                                                                          %

\bibitem{njp}
M.~Baiesi and C.~Maes, An update on nonequilibrium linear response. New J. Phys. {\bf 15}, 013004 (2013).

\bibitem{shuts}
E.~Levine, D.~Mukamel, and G.~M.~Sch\"utz, 
Zero-Range Process with Open Boundaries.
J. Stat. Phys. {\bf 120}, 759--778 (2005).



\bibitem{EH} 
M.~R.~Evans and T.~Haney, Nonequilibrium Statistical Mechanics of the Zero-Range Process and Related Models.
J. Phys. A: Math. Gen. {\bf 38} R195--R239 (2005).

\bibitem{bam09}
M.~Baiesi, C.~Maes and B.~Wynants, Fluctuations and response of nonequilibrium states. Phys. Rev. Lett.
{\bf 103}, 010602 (2009).


\bibitem{bmw1}
M.~Baiesi, C.~Maes and B.~Wynants, Nonequilibrium linear response for Markov dynamics, I: Jump 
processes and overdamped diffusions. J. Stat. Phys. \textbf{137}, 1094--1116 (2009).

\bibitem{mnw08}
C.~Maes, K.~ Neto\v{c}n\'{y}, and B.~Wynants,  On and beyond entropy production; the case of Markov jump
processes. Markov Proc. Rel. Fields 14, 445--464 (2008).


\bibitem{diez}
G.~Diezemann, Fluctuation-dissipation relations for Markov processes. Phys. Rev. E {\bf 72}, 011104 (2005).

\bibitem{mprf}
C.~Maes, and B.~Wynants,  On a response function and its interpretation. Markov Proc. Rel. Fields {\bf 16}, 45--58 (2010).


\bibitem{kubo66}
R.~Kubo, The fluctuation-dissipation theorem. Rep. Prog. Phys. {\bf 29}, 255---284 (1966).



\bibitem{gaw}
R.~Chetrite, G.~Falkovich, and K.~Gawedzki: Fluctuation relations in simple examples of non-equilibrium
steady states. J. Stat. Mech. P08005 (2008).


\bibitem{yan10} F.~Yang, Y.~Chen and Y.~Liu, The Green-Kubo formula for general Markov
processes with a continuous time parameter, J. Phys. A: Math. Theor. \textbf{43} 245002 (2010).

\bibitem{lor05}H.~A.~Lorentz, The motion of electrons in metallic bodies, Proc. Amst. Acad. \textbf{7}, 438 (1905).

\bibitem{dru00}P.~Drude, Zur Elektronentheorie der Metalle, Ann. Phys. 1, 566-612 (1900).

\bibitem{sza00} {\it Hard Ball Systems and the Lorentz Gas}, edited by 
D.~Szasz, Springer-Verlag, Berlin (2000). 

\bibitem{kla07} R.~Klages, {\it Microscopic Chaos, Fractals 
and Transport in Nonequilibrium Statistical Mechanics}, World Scientific, 
Singapore (2007).



\bibitem{splg03}H.~Larralde, F.~Leyvraz and C.~Mej\'{\i}a Monasterio, Transport properties of a modified Lorentz gas, 
J. Stat. Phys. \textbf{113}, Nos. 1/2,0197 (2003). 


\end{thebibliography}
\end{document}